\documentclass{article}

\usepackage{PRIMEarxiv}
\usepackage{float}
\usepackage[utf8]{inputenc} 
\usepackage[T1]{fontenc}    
\usepackage{hyperref}       
\usepackage{url}            
\usepackage{amsmath}
\usepackage{booktabs}       
\usepackage{amsfonts}       
\usepackage{nicefrac}       
\usepackage{microtype}      
\usepackage{lipsum}
\usepackage{fancyhdr}       
\usepackage{graph icx}       
\usepackage{authblk}
\graphicspath{{media/}}     

\title{Uncovering high-dimensional phase space and the application of Mixture of Experts (MoE) on building the Large CALPHAD Model (LCM)}

\author[1]{Zhengdi Liu}
\author[1]{Wenwen Sun\thanks{Corresponding author at: Department of Materials Science and Engineering,
		Southeast University, Nanjing 211189, China
		Email Address: swwcsu@live.cn}\textsuperscript{*,}}

\affil[1]{Jiangsu Key Laboratory for Advanced Metallic Materials, School of Materials Science and Engineering, Southeast University, Nanjing 211189, China}

\begin{document}
	\maketitle
	
	\begin{abstract}
		This study presents a novel approach for analyzing and establishing Large CALPHAD model (LCM) in complex alloy systems. Through the introduction of "composition space volume", a multi-dimensional metric which allows to quatitatively define alloy composition variations. Utilizing stochastic methods, the study quantifies phase space complexity through phase density, and model training costs through data density. This leads to a strategic segmentation of the entire composition space, tailored to the complexity of each segment, thereby reducing computational efforts in model training. A significant advancement is the integration of segmented models using a Mixture of Experts (MoE) approach, ensuring accurate portrayal of phase behaviors across diverse composition spaces. This technique is demonstrated in establishing a high-dimensional phase diagram for the FeCoNiTi system, highlighting its efficiency and accuracy. The study's methodologies offer a systematic and cost-effective framework for modeling complex alloy systems, marking a step forward in the field of alloy design and analysis.
	\end{abstract}


	\section{Introduction}
	The pursuit of advanced metal materials has progressively moved towards the exploration of Complex Concentrated Alloys (CCAs) and multi-principal element alloys (MPEAs), which promise to revolutionize our capabilities in materials design and engineering [1,2]. This trend is characterized by an increasing complexity in alloy systems, necessitated by the demand for materials with exceptional performance across a wide range of applications, including irradiation[3], corrosion-resistant[4], wear-resistant[5], electrical property[6], and superconducting materials development[7]. The rapid development of High-Entropy Alloys (HEAs) exemplifies the potential of these materials, revealing a wealth of properties that could not be achieved through traditional alloying strategies[8–10]. Despite their potential, a significant challenge that arises with CCAs is their designability. Traditional design methodologies that rely heavily on binary or ternary phase diagrams[11,12]. However, given their inherently limited scope in representing the complex interdependencies between composition and temperature in multi-element systems, they are less logical and detailed for the design of CCAs [13,14].
	
	To solve this problem, the design by high-throughput calculation based on thermodynamics is gradually developed. The efforts mainly focus on two parts, algorithms to search for certain phase transition or phase space [15,16], and the machine learning model to accelerate the thermodynamics calculation[17,18]. While they make progress, these methods relinquish the advantages of phase diagram-a panoramic perspective to design alloy. While obstruct by this dimension in real world, the advent of computational techniques has enabled the establishment of high-dimensional data, which can overcome the spatial limitations of traditional phase diagrams[19,20]. This is exemplified in the work by Liu et al., where the Large CALPHAD Model (LCM) was utilized to tackle the computational intensity of high-throughput calculations in the FeCrNiMn system, and paving the way for the establishment of high-dimensional phase diagram for alloy design in a panoramic perspective[21].
	
	However, the approach of randomly generating data points across the phase space for dataset creation, as utilized in Liu et al.'s study, operates under the assumption that training the LCM (LCM) incurs a uniform cost for every part of the phase space. This method potentially overlooks the varying complexities inherent within different regions of the phase space. Empirical observations suggest that phase diagram complexity varies across different compositional regions, which is often due to the formation of specific compounds[22,23] . This variability becomes even more pronounced as the number of elements in the alloy system increases, extending the phase space into higher dimensions and exacerbating the challenge of accurate representation and analysis[24,25].
	
	To address this, the present work proposes a set of definitions and quantitative equations for assessing phase space complexity. Through the application of ergodic and stochastic methods, we offer a comprehensive framework for understanding the intricate behaviors within multi-dimensional phase spaces.
	
	Building on this theoretical foundation, the study proceeds to qualitatively analyze the training costs associated with modeling phase spaces of varying complexities. It becomes evident that a one-size-fits-all approach is suboptimal. Instead, a strategic segmentation of the entire composition space is proposed to enhance the efficiency and accuracy of LCM training, with the potential for significant cost reductions.
	
	Each segmented composition space is then modeled to develop part-specific LCMs[26]. These models are integrated using a Mixture of Experts (MoE) approach, ensuring a seamless transition between different phases and complexities. This integration is not only a technical advancement but a strategic innovation that enables the accurate and efficient development of LCMs for increasingly complex alloy systems without the prohibitive costs typically associated with high-dimensional modeling.
	
	Finally, we explore the establishment of a high-dimensional phase diagram for the FeCoNiTi system, drawing parallels to Liu et al.’s methodology while highlighting the novel contributions of this study. The detailed distribution of data points across various phase statuses will be presented, emphasizing the robust data foundation available for alloy design[21].

	\section{Methods}
	\label{sec:headings}
	
	\subsection{Composition space and its volume}
	In the study of multi-principle elements alloy systems, the concept of "composition space" is crucial. The composition space is a multi-dimensional construct where each axis represents the concentration of a particular element in the alloy. Within this space, "composition space volume" is a defined, measurable region that encapsulates the range of possible alloy compositions.
	
	Defining composition space volume:
	
	1. For a binary system: The composition space is represented as a line segment along one axis, typically scaled from 0 to 100\% for one component. The line illustrates the varying proportions of the two components in the alloy. The length of this line segment denotes the composition space volume.
	
	2. For a ternary system: In this case, the entire composition space forms an area of an equilateral triangle, with each vertex representing a pure component. The area of this triangle constitutes the composition space volume, representing all possible combinations of these three components.
	
	3. For a quaternary system and beyond: In systems with four or more components, the composition space becomes a regular tetrahedron or higher-dimensional simplices, respectively. The volume of these geometric shapes represents the composition space volume, encompassing the full range of compositional variations.
	
	Principles governing the geometry of composition space:
	
	1. Elemental equivalence: This principle ensures that each element contributes equally to the composition space. It maintains equidistance between elements in the geometric representation, ensuring a fair representation of each component.
	
	2. Uniform geometric representation: The composition space retains a consistent geometric shape across different alloy systems. This uniformity aids in comparative analysis, making it easier to understand and explore the phase behavior and properties across various alloy compositions.
	
	The advantage of this definition is illustrated by a ternary system example. As shown in Fig. 1a and 1b, two red triangles represent the components A range from 20\% to 50\%, B range from 20\% to 50\%, and C range from 30\% to 60\%. When we only concentrate on the phase status within a certain component area, both these equilateral triangle and isosceles right triangle can represent the phase status well. But when we want to define the composition space volume, some problem occurs. In Fig. 1b, the components were represented in an isosceles right triangle. the three side lines represent the same component range but have different geometric length. If we use A and C as the representation of this area, the area can be defined with these conditions:
	
	\begin{equation}
		\left\{
		\begin{array}{l}
			20 \leq x(A) \leq 50 \\
			30 \leq x(C) \leq 60 \\
			x(A) + x(C) \leq 80
		\end{array}
		\right.
	\end{equation}
	
	And the composition space volume can be calculated as the area of the isosceles right triangle, which is 30*30/2 = 450. Under this situation, the geometric representation and component volume has a reasonable correspondence. However, when we set this area by A and B, the area is defined by these conditions:
	
	\begin{equation}
		\left\{
		\begin{array}{l}
			20 \leq x(A) \leq 50 \\
			20 \leq x(B) \leq 50 \\
			x(A) + x(B) \leq 70
		\end{array}
		\right.
	\end{equation}
	
	while this geometric volume contradictory to the geometric volume when using A and C define this area. But the geometric volume is equivalent when using equilateral triangle, so the composition space volume is better to define with the geometric volume in an equilateral triangle.
	
	\begin{figure}
		\centering
		\includegraphics[width=0.8\textwidth]{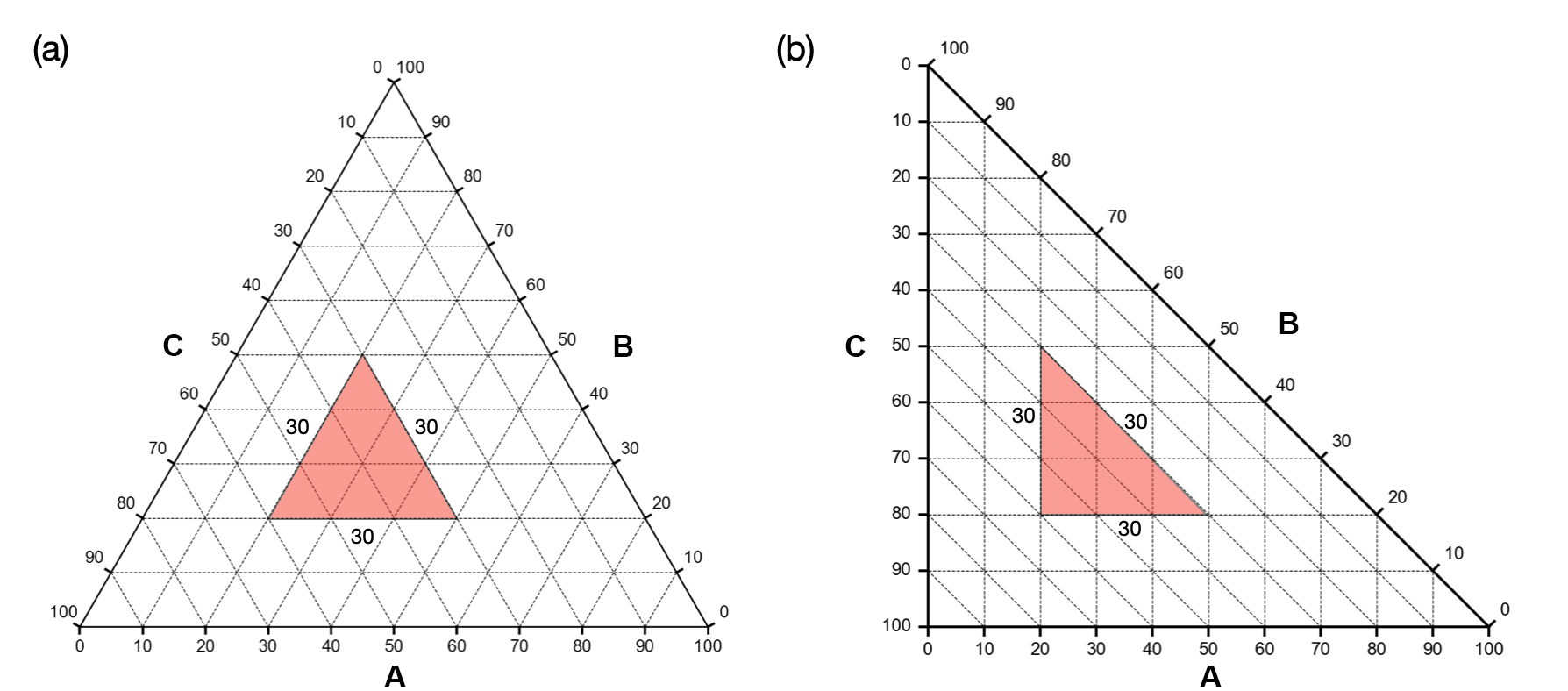}
		\caption{(a) Using equilateral triangle to represent composition space in ternary system. (b) Using isosceles right triangle to represent composition space in ternary system.}
		\label{fig:fig1}
	\end{figure}
	
	Promote this property to n elements system. It is better to define the composition space volume in a n-1 regular simplex. Table. 1 illustrates the total volume of 3 to 6 elements system when the side length is set equal to 100. For a j dimension regular simplex, the equation to calculate volume is defined as [27]:
	\begin{equation}
		V_j^2(S) = \frac{(-1)^{j+1}}{2^j (j!)^2} \det(\hat{B})
	\end{equation}
	Where B is the (j+2) * (j+2) matrix:
	\begin{equation}
		B=
		\begin{bmatrix}
			0&  d_{01}^{2} &  d_{02}^{2}&  \dots &  d_{0j}^{2}&1 \\
			d_{10}^{2}&  0&  d_{12}^{2}& \dots &  d_{1j}^{2}&1 \\
			d_{20}^{2}&  d_{21}^{2}&  0&  \dots&  d_{2j}^{2}&1 \\
			\vdots &  \vdots&  \vdots&  \ddots&  \vdots&\vdots \\
			d_{j0}^{2}&  d_{j1}^{2}&  d_{j2}^{2}&  \dots&  0&1 \\
			1&  1&  1&  \dots&  1&0
		\end{bmatrix}
	\end{equation}
	
	Where d$_{ij}$ is the distance between i vertex and j vertex. For a regular simplex, all d$_{ij}$s have the same value. Table. 1 shows the alloy systems with different number of elements and their corresponding composition space volume.
	
	\begin{table}[ht]
		\centering
		\caption{The composition space volume for alloy system with different number of elements.}
		\label{tab:composition-space-volume}
		\begin{tabular}{ccc}
			\toprule
			Number of elements & Dimension & Composition space volume \\
			\midrule
			3 & 2 & \( 2500\sqrt{3} \) \\
			4 & 3 & \( \frac{250000\sqrt{2}}{3} \) \\
			5 & 4 & \( \frac{10^8 \sqrt{5}}{96} \) \\
			6 & 5 & \( \frac{10^9 \sqrt{3}}{48} \) \\
			\bottomrule
		\end{tabular}
	\end{table}
	
	\subsection{Number of phase spaces and phase space complexity.}
	
	\subsubsection{Ergodic method}
	The ergodic method is employed to systematically analyze the complexity of phase space in an alloy system. This method involves an exhaustive exploration of the phase space, which is defined by number of different phase statuses at various compositions and temperatures.
	
	The FeNiTi System is used as an example in this study. For the FeNiTi system, the ergodic method involves calculating phase statuses at increment of 1\% in composition and 20K in temperature. This detailed approach allows to investigate the entire phase space. The analysis encompassed 391,476 points, requiring approximately 39 hours to complete. This duration is manageable for a ternary system like FeNiTi.
	
	The analysis reveals significant variability in phase space complexity across different compositional areas, as illustrated in Fig. 2a. An area of increased complexity, marked in the red block of Fig. 2b, is identified where the Ni to Ti ratio is around 3:1. This complexity is primarily attributed to the formation of the Ni3Ti compound. The phase diagram in Fig. 2c, corresponding to the yellow line in Fig. 2b, highlights the region (where x(Fe) $\approx$ 15\%) with the highest complexity within the FeNiTi phase space.
	
	While the ergodic method is viable for a ternary system, scaling it to systems with four or more elements introduces significant challenges. These include the increased complexity in tracking phase behavior across a larger composition space and the exponential growth in computational time. For a quaternary system, the analysis could extend to several months.
	Due to these challenges, particularly for higher-dimensional systems, we advocate for the use of a stochastic method. This approach aims to efficiently characterize phase space complexity without the extensive computational demands associated with the ergodic method.
	
	\begin{figure}
		\centering
		\includegraphics[width=1\textwidth]{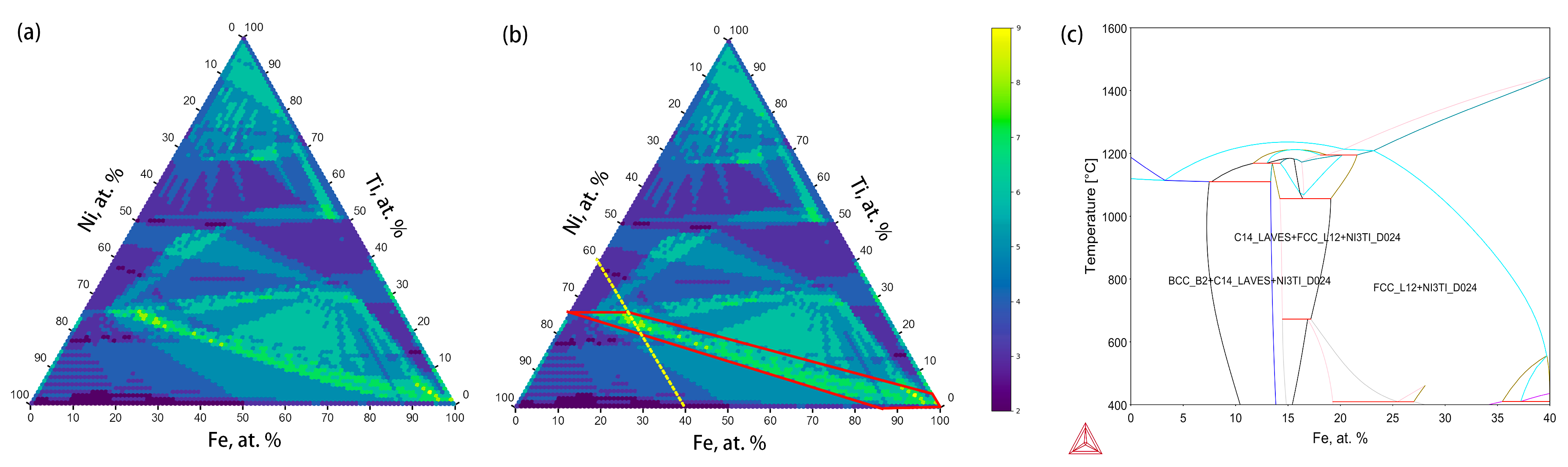}
		\caption{(a) The number of phase statuses contains in each composition of FeNiTi system. (b) The red block is the area where the Ni to Ti ratio is around 3:1. (c) The phase diagram which the yellow line in Fig. 2b represented, it penetrate the most complex area in FeNiTi system.}
		\label{fig:fig2}
	\end{figure}
	
	\subsubsection{Stochastic method for high-dimensional alloy systems}
	The application of the ergodic method for phase space analysis, while comprehensive, becomes impractical for high-dimensional systems such as quaternary or more elements alloy systems due to the immense time and computational resources required. To address this, a stochastic method is proposed. This approach efficiently assesses phase space complexity to balance data comprehensiveness with practical feasibility.
	
	1. Framework of the stochastic method
	
	In the stochastic method, the primary goal is to identify and quantify the non-uniform distribution of phase spaces within the composition space of the alloy system.
	
	As shown in Fig. 3a, to attain the number of phase spaces, there are total four steps.
	
	$\bullet$ Generate random data: Random data points are generated within the phase space. The objective is to collect sufficient data to discern the non-uniformity in phase distributions.
	
	$\bullet$ Choose composition space domain: The composition space domain is chosen as a regular n-simplex, which is used to scan the entire phase space. This selection helps to systematically cover the whole phase space.
	
	$\bullet$ Scan the phase space: The phase space is methodically scanned using the composition space domain, with step lengths adjusted to ensure thorough coverage.
	
	$\bullet$ Data analysis: Each step involves analyzing the random data within the composition space domain to identify variations in phase space complexity.
	
	2. Application to the FeCoNiTi system
	
	For the FeCoNiTi system, a regular tetrahedron with a 20-unit side length is used as the composition space domain, and the step is set at 10. The analysis revealed significant variations in the complexity of phase spaces, indicating a high degree of non-uniformity in the phase space. 
	
	In Fig. 3b, the relationship between the cumulative number of phase spaces (Np) identified in each scanned domain and the total number of random data points is analyzed. Initially, as the number of data points increases from 0 to 25,000, there is a dramatic rise in the sum of identified phase spaces. This steep increase suggests that many phase spaces were initially overlooked, reflecting the under-representation of phase space complexity with fewer data points. However, beyond 100,000 data points, this sum begins to stabilize, indicating a convergence toward a more accurate representation of the phase space complexity within the system.
	
	For a more intuitive analysis of these complex multi-dimensional spaces, we implemented dimensionality reduction techniques, as illustrated in Fig. 3c. Each point in this reduced space, such as one with coordinates (10, 20, 20), represents a specific scanned domain within the composition space, offering insights into the localized complexity of phase spaces. For instance, this particular point corresponds to a domain defined as follows:
	\begin{equation}
		\left\{
		\begin{array}{l}
			10 \leq x(\text{Fe}) \leq 30 \\
			20 \leq x(\text{Co}) \leq 40 \\
			20 \leq x(\text{Ni}) \leq 40 \\
			x(\text{Fe}) + x(\text{Co}) + x(\text{Ni}) \leq 70
		\end{array}
		\right.
	\end{equation}
	
	This scanning method reveals significant insights into the phase space complexity. At a lower data point threshold in Fig. 3c, which represents 3,000 data points, the number of phase spaces detected is noticeably underestimated, indicating an incomplete representation of phase space complexity. However, as the data points increase to 30,000, as shown in Fig. 3d, a clearer trend in phase space complexity starts to emerge, albeit without precise detail. This trend becomes more distinct and informative at higher data points, as shown in Fig. 3e and f representing 80,000 and 150,000 data points. The similarity in the complexity patterns observed at these two higher data thresholds suggests that 80,000 data points are sufficient for accurately distinguishing the complexity within the phase space, although 150,000 points provide a more detailed view.
	
	Nonetheless, from a practical standpoint, the analysis with 80,000 data points effectively balances detail with computational efficiency. It provides a reliable understanding of the phase space complexity, which is essential for the subsequent training of the LCM. This balance is critical, considering that the cost associated with training an LCM is significantly higher than that required for distinguishing phase space complexity.
	
	\begin{figure}
		\centering
		\includegraphics[width=1\textwidth]{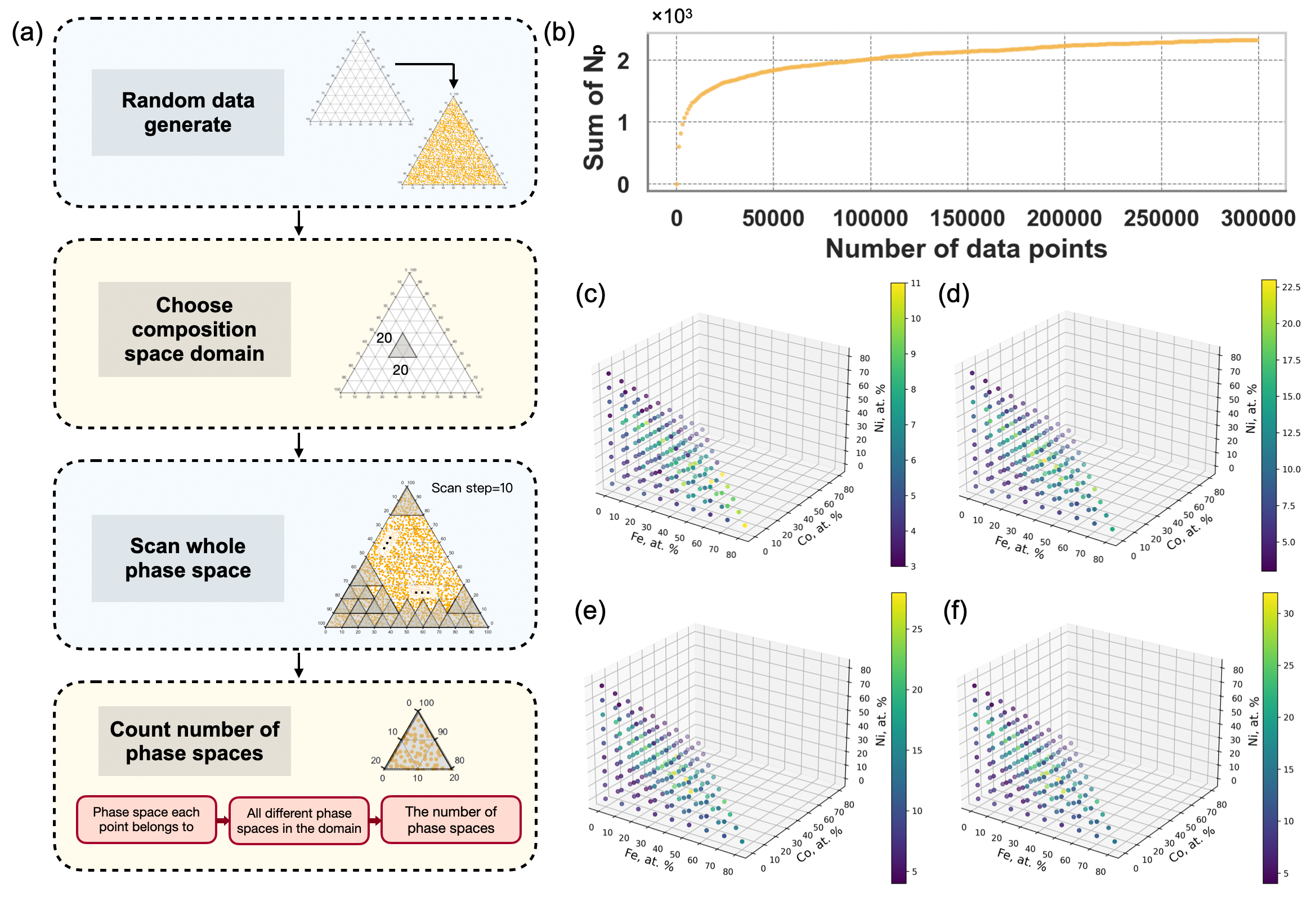}
		\caption{(a) The framework of the stochastic method to evaluate the phase space complexity. (b) The relationship between number of data points and the sum of number of phase space in each scan composition space domain. (c-f) The visualization of the distribution of phase space complexity with 3000, 30000, 80000, and 150000 data points, respectively.}
		\label{fig:fig3}
	\end{figure}
	
	\subsection{Relationship between phase space complexity and training costs}
	As above mentioned, in FeCoNiTi system, the phase space complexity is not uniform throughout the phase space, which will lead to varied costs in achieving an accurate model. Therefore, there is a need for a quantitative approach to describe the complexity of the phase space and the corresponding training costs accurately. 
	
	In the subsequent analysis, phase density will be employed to quantify the complexity within specific composition space. Simultaneously, data density will be utilized to gauge the cost of training the model to the desired level of accuracy. Together, these metrics will facilitate a more systematic and cost-effective approach to modeling complex alloy systems.
	
	\subsubsection{Quantifying complexity through phase density}
	Phase density (D$_p$) is defined as the ratio of the number of distinct phase spaces (N$_p$) to the composition space volume of the composition space being examined (V), expressed as:
	\begin{equation}
		D_{p} = \frac{N_{p}}{V}
	\end{equation}
	
	High phase density signifies regions within the composition space that have a large number of phase space, requiring more data points to accurately capture and model.
	
	\subsubsection{Data density in model training}
	The concept of data density (D$_d$), which represents the number of data points per unit composition space volume (V), express as:
	\begin{equation}
		D_{d} = \frac{N_{d}}{V}
	\end{equation}
	
	Where N$_d$ is the number of data points in examined composition space. It emerges as a critical determinant of precision and cost of the modeling process. A higher data density may be required to resolve the complex phase behaviors associated with high phase densities.
	
	\subsubsection{Correlating data density with training cost}
	The study focuses on nine composition spaces, each with a side length of 20 units. These composition spaces were selected based on varying numbers of phase spaces. Table. 2 lists the start points of these composition spaces and the corresponding number of phase spaces and phase densities.
	
	\begin{table}[ht]
		\centering
		\caption{Characteristics of selected composition Spaces in the FeCoNiTi system.}
		\label{tab:my_label}
		\begin{tabular}{cccccc}
			\toprule
			\multicolumn{4}{c}{\textbf{Start point}} & \textbf{Number of phase spaces} & \textbf{Phase density} \\
			\cmidrule(lr){1-4} 
			\textbf{Fe, at. \%} & \textbf{Co, at. \%} & \textbf{Ni, at. \%} & \textbf{Ti, at. \%} & & \\
			\midrule
			20 & 0 & 5 & 55 & 6 & \(6.36\times10^{-3}\) \\
			20 & 5 & 0 & 55 & 7 & \(7.42\times10^{-3}\) \\
			20 & 10 & 10 & 40 & 9 & \(9.55\times10^{-3}\) \\
			20 & 15 & 10 & 35 & 12 & \(1.27\times10^{-2}\) \\
			20 & 20 & 10 & 30 & 16 & \(1.70\times10^{-2}\) \\
			20 & 25 & 5 & 30 & 20 & \(2.12\times10^{-2}\) \\
			20 & 25 & 15 & 20 & 25 & \(2.65\times10^{-2}\) \\
			20 & 40 & 0 & 20 & 35 & \(3.71\times10^{-2}\) \\
			20 & 35 & 0 & 25 & 39 & \(4.14\times10^{-2}\) \\
			\bottomrule
		\end{tabular}
	\end{table}
	
	Fig. 4a presents a comparative analysis of nine different phase densities in the FeCoNiTi system, each represented by a distinct line. These lines trace the relationship between data density and the mean squared error (MSE) of the trained model, demonstrating how the phase density affects the data requirement for accurate modeling.
	
	For lines indicating lower phase densities (e.g., the blue line with a phase density of 6.36$\times10^{-3}$), the MSE is relatively low even at smaller data densities. This suggests that when the complexity of phase space is minimal, a smaller data density is sufficient to train an accurate model. In contrast, lines representing higher phase densities (such as the orange line with a phase density of 4.14$\times10^{-2}$) show that a greater data density is necessary to achieve a comparable MSE. This implies that as the phase space complexity increases, more data points are required to accurately model the system.
	
	Notably, the line with a phase density of 1.27$\times10^{-2}$ (purple line with huge fluctuation) demonstrates a phenomenon: as data density increases, there is an initial rise in MSE. This initial rise in MSE suggests that the model, when provided with insufficient data relative to the complexity of the phase space, may not fully account for all the significant features. However, as data density continues to increase, the model begins to integrate these complexities more accurately, which is reflected in the decreasing MSE.
	
	To ensure the reliability of the LCM in alloy design and the construction of high-dimensional phase diagrams, an accuracy baseline has been established at an MSE of 1$\times10^{-4}$. Models that achieve an MSE below this threshold are considered dependable for practical applications in alloy design. This criterion serves as a quantitative benchmark to ascertain the fidelity of predictions of the model, thereby facilitating the development of materials with tailored properties and the accurate mapping of complex phase spaces.
	
	The above analysis focuses on composition space domains with a side length of 20 units. The follow investigation delves into how training costs correlate with the size of the composition space, particularly when these spaces exhibit similar phase densities. Table. 3 presents the details of the composition space domains under study, summarizing their positions and associated phase densities.
	
	\begin{table}[ht]
		\centering
		\caption{Composition space domains information used to investigate the relationship between size of the composition space and training cost.}
		\label{tab:my_label}
		\begin{tabular}{ccccccc}
			\toprule
			\multicolumn{4}{c}{\textbf{Start point}} & \textbf{side length} & \textbf{Number of phase spaces} & \textbf{Phase density}  \\
			\cmidrule(lr){1-4} 
			\textbf{Fe, at. \%} & \textbf{Co, at. \%} & \textbf{Ni, at. \%} & \textbf{Ti, at. \%} & & \\
			\midrule
			20 & 0 & 5 & 55 & 20 & 6 & \(6.36\times10^{-3}\) \\
			20 & 5 & 0 & 55 & 25 & 12 & \(6.52\times10^{-3}\) \\
			20 & 10 & 10 & 40 & 30 & 21 & \(6.59\times10^{-3}\) \\
			\bottomrule
		\end{tabular}
	\end{table}
	
	Fig. 4b delineates the correlation between the MSE and data density within composition spaces of various side lengths, each exhibiting a similar phase density. This similarity in phase density suggests comparable levels of complexity across the composition spaces analyzed.
	
	The rationale behind this trend is that in larger composition spaces, the probability of completely capturing the boundaries of individual phase spaces increases. This comprehensive inclusion means that all data points that define a particular phase space are more likely to be contained within the composition space. Therefore, as the composition space expands, it naturally integrates more whole phase spaces and the data points describing them, providing a more complete and detailed dataset for the model to learn from.
	
	This observation leads to a strategic insight for model training: when phase density is consistent, selecting a larger composition space is beneficial. It allows for a more thorough representation of the phase behavior, which is crucial for the accuracy of the model. This approach ensures that the intricate details of the phase transitions and the characteristics of the phase spaces are well-represented, enhancing the model's ability to make precise predictions while optimizing the efficiency of data utilization.
	
	\begin{figure}
		\centering
		\includegraphics[width=1\textwidth]{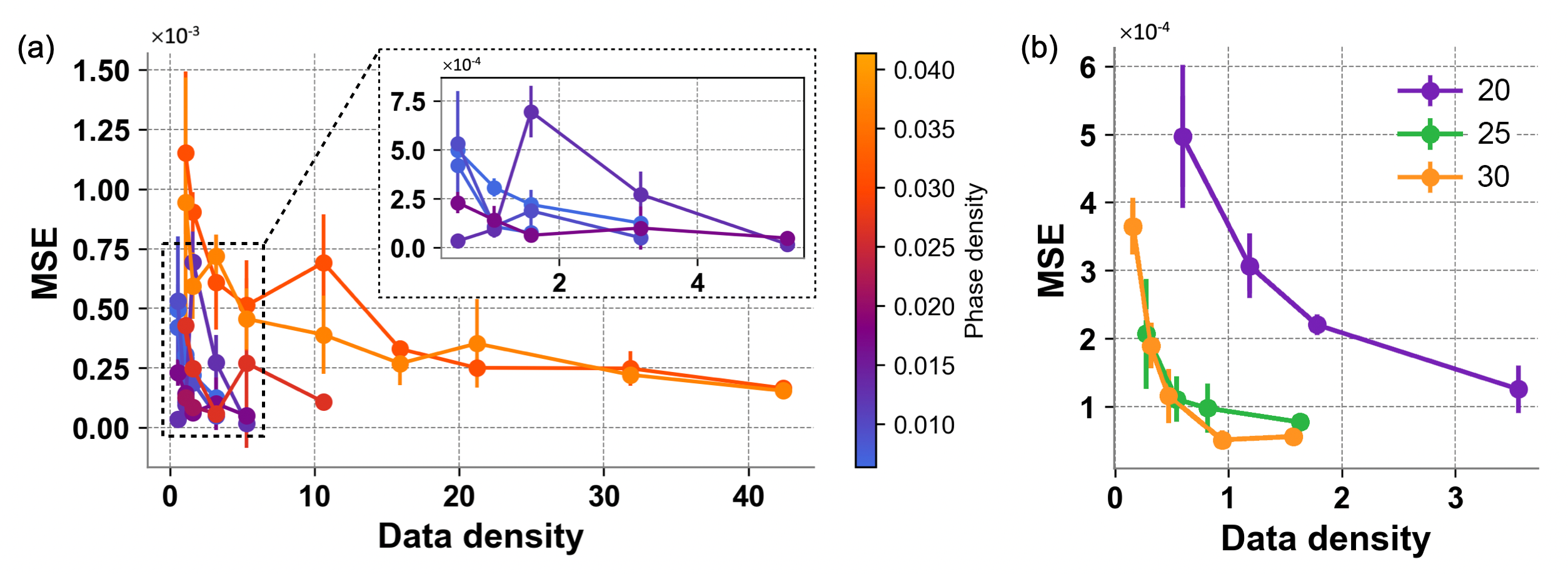}
		\caption{(a) The relationship between data density and MSE of nine investigated composition spaces with different phase density. (b) the relationship between data density and MSE of different composition spaces with different side length but similar phase density.}
		\label{fig:fig4}
	\end{figure}
	
	\section{Result and Discussion}
	\subsection{A systematic way to divide the composition space}
	With a foundational analysis of relationship between phase density, data density, and the volume of composition space domains complete, the focus now shifts to developing a systematic method for dividing the composition space. This division is pivotal not just for maintaining the high accuracy of predictive models but also for minimizing the costs associated with their training. As the complexity of alloy systems increases, encompassing alloys with five or six elements or steels with intricate carbide formations, the cost of developing predictive models can increase rapidly. However, by employing a methodical division strategy to divide the composition space, these costs can be managed effectively. This approach not only refines the modeling process but also ensures that the precision of alloy design and analysis remains high without incurring prohibitive costs. It is through such strategic division that the development of LCMs for these complex systems remains feasible and economically viable.
	
	Criteria for systematic division
	The exploration into the relationship between phase density, data density, and composition space domain volume has yielded two key qualitative insights:
	
	$\bullet$ Disparate phase density implications: Significant disparities in phase density correlate with increased costs in model training. This variance necessitates more data to accurately capture the complex phase behaviors, leading to higher computational expenses.
	
	$\bullet$ Size efficiency for comparable phase densities: When phase densities are similar, opting for a larger continuous composition space can effectively reduce the cost of training. A more extensive continuous space tends to include more complete phase information, which can enhance the model's learning process with the need for less data.
	
	From these insights, two principles for dividing the composition space have been derived:
	
	$\bullet$ Differential treatment based on phase density: Composition spaces exhibiting varying phase densities should be segmented and approached differently in the model training process. This differentiation ensures that each space is allocated resources proportionate to its complexity.
	
	$\bullet$ Maximizing composition space for uniform phase densities: In cases where phase densities are similar, efforts should be made to delineate as large a continuous composition space as is feasible. The aim is to encompass a comprehensive set of phase information within a single domain, thereby streamlining the data requirements and reducing training costs.
	
	In the presented study, the entire composition space of the FeCoNiTi system is systematically divided according to phase density categories: low (D$_p$ < 2.33$\times10^{-2}$), medium (2.33$\times10^{-2}$ $\le$ D$_p$ < 3.18$\times10^{-2}$), and high (3.18$\times10^{-2}$ $\le$ D$_p$). To maintain the continuity of each divided space, the Depth-First Search (DFS) algorithm is utilized. 
	
	Fig. 5a displays the outcome of this division, showcasing three distinct, continuous composition spaces that correspond to the predefined phase density criteria. It should be noted that while the division results have been visualized for a quaternary system, the methodology employed is straightforward and adaptable for delineating the composition space in higher-dimensional systems, including those with five or more elements. The approach's simplicity and clarity are maintained even as the number of elements—and thus the complexity—increases.
	
	Fig. 5b illustrates the volume of composition space allocated to each divided segment based on phase density. It is observed that the segment categorized as low phase density encompasses the largest portion, accounting for 97.2\% of the entire composition space. Conversely, the high phase density segment constitutes a mere 6.3\% of the total composition space. This discrepancy in volume distribution is a direct result of the phase density determination method, which relies on scanning distinct composition space domains. Due to this scanning approach, there is an inherent overlap between segments, leading to a cumulative composition space volume that exceeds the volume of the whole space.
	
	Fig. 5c depicts the relationship between data density and MSE for composition spaces classified by low, medium, and high phase density. It is evident that the largest composition space, with low phase density, requires the least data density to meet the established accuracy baseline. In contrast, the smallest composition space, with high phase density, demands almost an order of magnitude greater data density to achieve the same level of accuracy. Table 4 details the data density necessary to train models for each phase density category to the accuracy baseline.
	
	\begin{table}[ht]
		\centering
		\caption{Training costs for composition spaces to reach accuracy baseline}
		\label{tab:training-costs}
		\begin{tabular}{ccc}
			\toprule
			Composition space & Data density & Number of data points \\
			\midrule
			\(D_p < 2.33 \times 10^{-2}\) & 5.24 & 600000 \\
			\(2.33 \times 10^{-2} \leq D_p < 3.18 \times 10^{-2}\) & 11.42 & 400000 \\
			\(3.18 \times 10^{-2} \leq D_p\) & 40.24 & 300000 \\
			\bottomrule
		\end{tabular}
	\end{table}
	
	Significantly, the total number of data points required to train models for the three different composition spaces amounts to approximately 1,050,000 (apart from overlaps). Without employing this methodical division, the whole composition space is treated uniformly—as if it possessed a uniformly high phase density. Such an indiscriminate approach would necessitate an overwhelming 4,740,000 data points.
	
	This method underscores the efficiency to training models within distinct composition spaces. By avoiding the assumption of uniform phase density across the entire space, a significant reduction in data requirements is achieved, demonstrating the cost-effectiveness of the systematic division strategy.
	
	\begin{figure}
		\centering
		\includegraphics[width=1\textwidth]{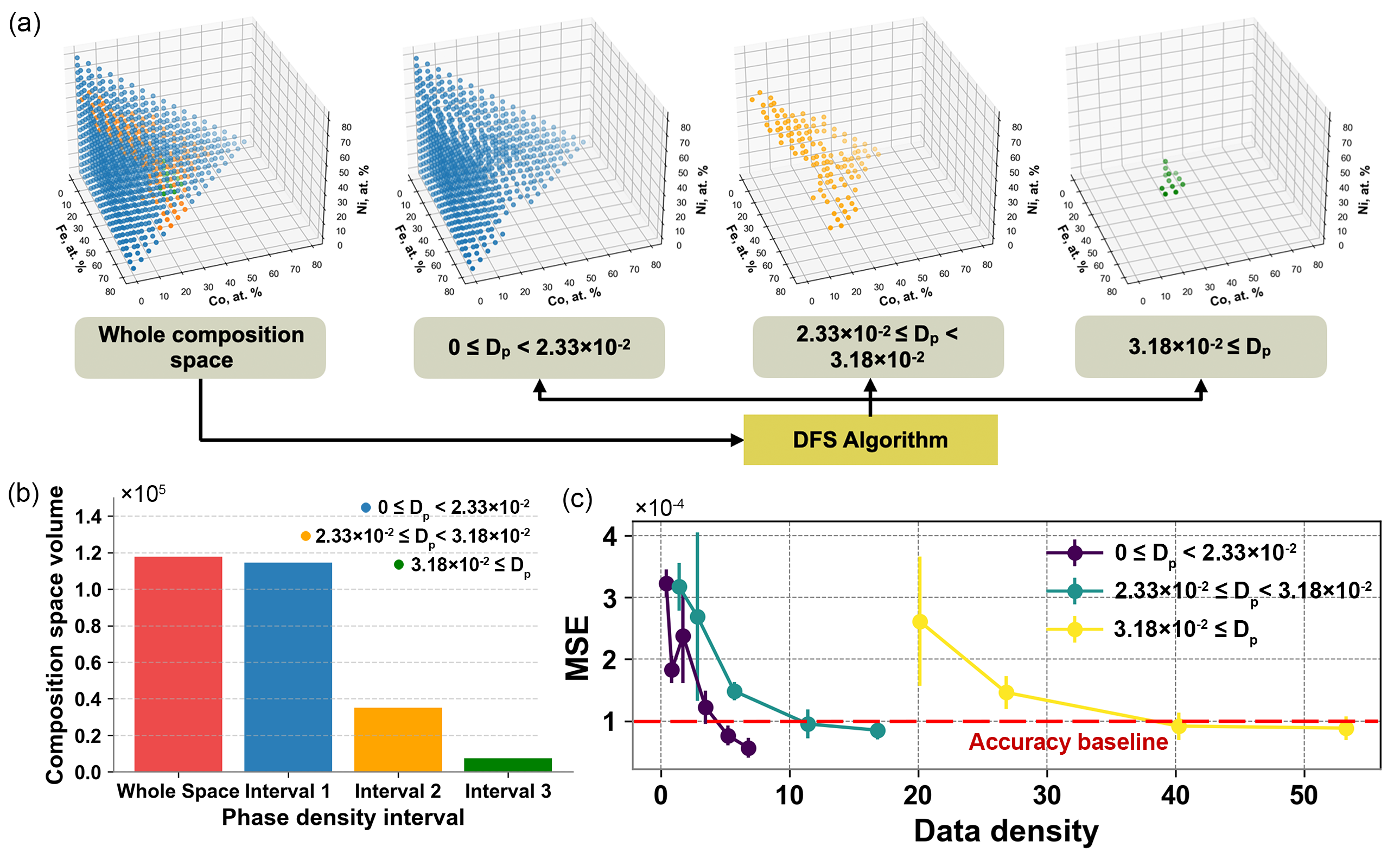}
		\caption{(a) DFS algorithm divides the whole composition space according to different phase densities. (b) The phase volume of the whole composition space and divided composition spaces. (c) The necessary data density to train the composition with different D$_p$ to accuracy baseline.}
		\label{fig:fig5}
	\end{figure}
	
	\subsection{Integration of trained models using Mixture of Experts (MoE)}
	Upon successfully training individual models for low, medium, and high phase density composition spaces, the next step is the synthesis of these models into a cohesive framework. This is accomplished through a MoE approach, as depicted in Fig. 6a.
	
	The MoE architecture is specifically designed to leverage the strengths of multiple specialized models[26]. The MoE model in this study is a sophisticated ensemble framework that integrates distinct predictive models, each trained on composition spaces with varying phase densities within the FeCoNiTi system. Inputs to the MoE model represent the full spectrum of the composition space, and through a probabilistic gating mechanism, data points are given a confidence level to each 'expert'. Each expert model, specialized for low, medium, or high phase densities, processes inputs to generate predictions. These outputs are then weighted by the gating mechanism's confidence levels, reflecting the relative expertise of each model for the given data. The collective outputs are aggregated and normalized to deliver the final MoE model prediction, which is a nuanced amalgamation of insights drawn from the entire composition space.
	
	The effectiveness of the MoE model is visually underscored by phase diagrams corresponding to a specific composition within the FeCoNiTi system, with x(Ni) at 0.01\% and x(Co) at 40\%. Fig. 6b presents three phase diagrams generated by three different experts, each responsible for a specific phase density range within the FeCoNiTi system. Expert 1 captures the phase behaviors in the low phase density composition space, while Expert 3 specializes in the high phase density regions, providing detailed insights.
	
	Fig. 6c showcases the integrated phase diagram produced by the MoE model, synthesizing the contributions from all experts. When compared with the phase diagram from the Thermo-Calc console depicted in Fig. 6d, the MoE model's output exhibits a high degree of similarity, indicating its accuracy in representing both high and low phase density composition spaces. This comparison underscores the MoE model's capability to accurately render complex phase behaviors across the entire composition space, validating its application as a robust tool in alloy design and analysis.
	
	In summary, the Mixture of Experts (MoE) model constitutes an advanced ensemble framework that not only leverages the specialized expertise of individual models across different phase density regions but also ensures a seamless transition between these experts. This continuity is crucial for the accurate portrayal of phase transitions and behaviors in alloy systems. Moreover, as the complexity of the alloy systems increases, with an accompanying rise in the intricacy of the phase space, the MoE model offers a streamlined solution. It enables the development of the LCM in complex alloy systems to be entirely data-driven, obviating the need for direct visualization or data validation, rendering it a potent and pragmatic tool in the realm of alloy design and analysis.
	
	\begin{figure}
		\centering
		\includegraphics[width=1\textwidth]{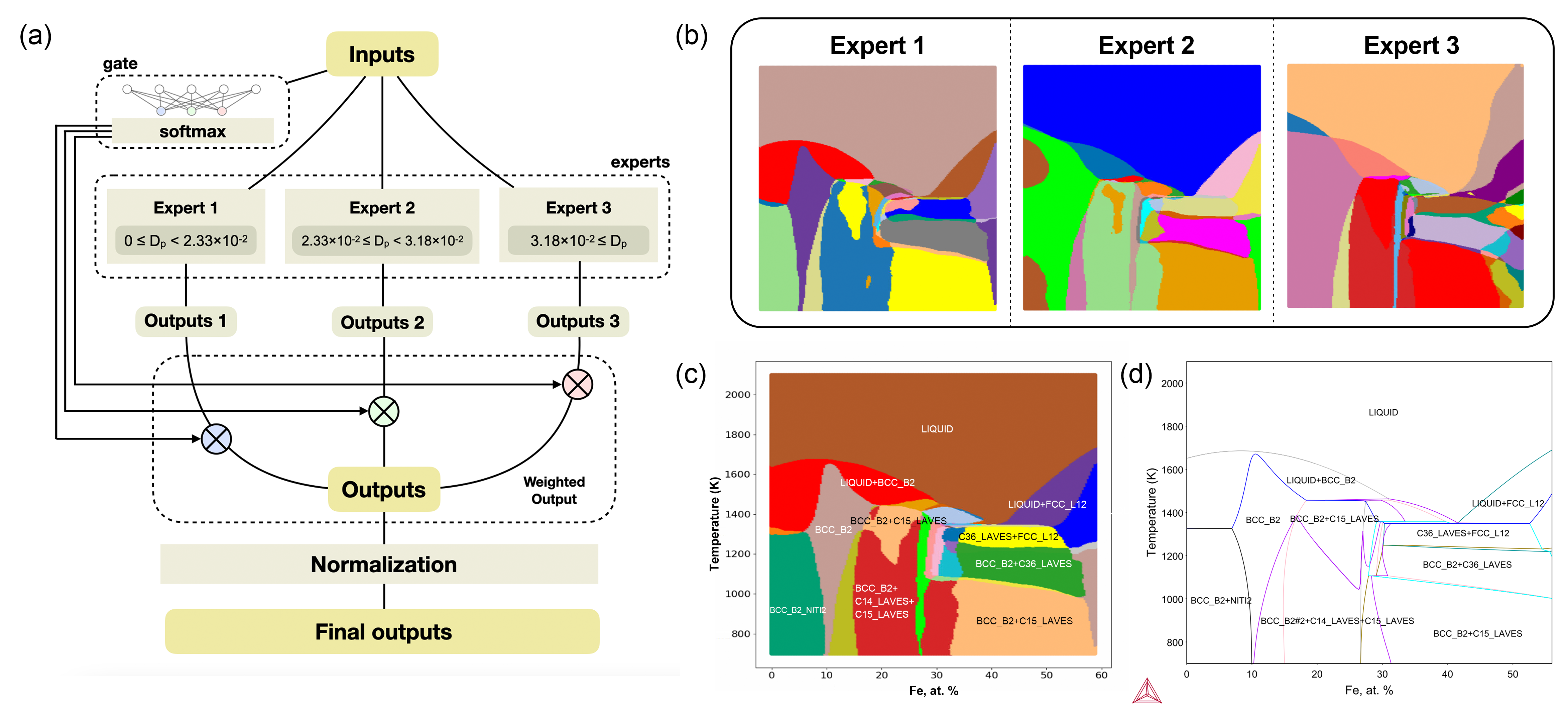}
		\caption{(a) The prediction method of MoE model. (b) Three phase diagrams drawn by different experts. (c) The phase diagram drawn by MoE model. (d) The phase diagram drawn by Thermo-Calc console. x(Co) is set at 40\% and x(Ni) is set at 0.01\%, and x(Fe) is changing from 0\% to 59\% for all phase diagrams.}
		\label{fig:fig6}
	\end{figure}
	
	\subsection{The establishment of high-dimensional phase diagram in FeCoNiTi system}
	The approach for constructing a high-dimensional phase diagram in the FeCoNiTi system aligns with the method outlined in Liu et al.'s work[21]. Fig. 7a outlines the framework utilized to establish the high-dimensional phase diagram. In the current study, the phase space is densely sampled with a composition increment of 1\% and a temperature increment of 7.5K, ranging from 673.15K to 2173.15K. This results in a total of 35,723,902 data points that form the comprehensive inputs for the phase space analysis.
	
	Following the calculation of these inputs, the resulting phase status data is organized into a hashed list for efficient retrieval. The DFS algorithm is then employed to delineate independent phase spaces, where any phase space containing more than 50 data points is deemed significant for alloy design purposes. Utilizing this criteria, the DFS algorithm identifies 95 distinct phase statuses within the FeCoNiTi system. Each identified phase status is subsequently stored separately, facilitating quick and easy access for future alloy design requirements.
	
	Fig. 7b presents the distribution of data points across the various phase statuses identified in the FeCoNiTi system, showcasing the data-rich foundation for the alloy design demand.
	
	\begin{figure}
		\centering
		\includegraphics[width=1\textwidth]{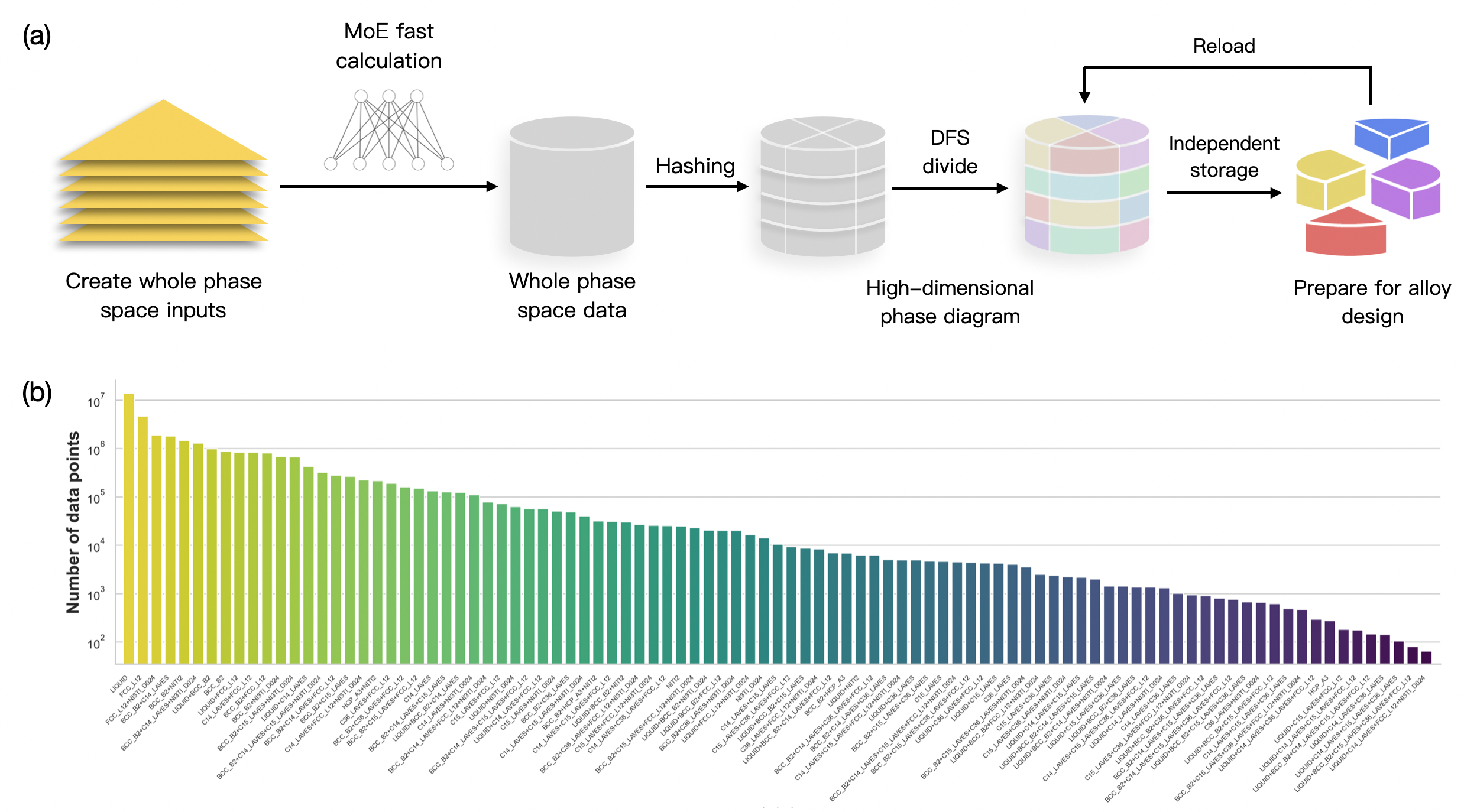}
		\caption{(a) The framework of establishing high-dimensional phase diagram. (b) The number of data points for each type of phase status in FeCoNiTi system.}
		\label{fig:fig7}
	\end{figure}
	
	\section{Conclusion}
	This study presents a comprehensive approach to understanding and modeling the complexity of high-dimensional phase spaces in CCAs and MPEAs, with a particular focus on the FeCoNiTi system. The novel concepts of "composition space" and its volume have been introduced, offering a new perspective on representing alloy compositions in multi-dimensional spaces. This framework not only aids in accurately defining and measuring the compositional range of alloys but also sets the foundation for analyzing phase space complexity through ergodic and stochastic methods.
	
	In this work, the quantitative assessment of phase space complexity using the newly defined metrics of phase density and data density is proposed. This assessment highlights the non-uniformity of phase space complexity and underscores the inefficiency of treating all regions of the phase space equally in terms of modeling cost. The study proposes a systematic segmentation of the composition space, based on phase density, to optimize the training of LCMs. This strategy not only ensures high accuracy in predictive models but also significantly reduces associated costs, particularly for complex alloy systems.
	
	The integration of individually trained models for different segments of the composition space using the MoE approach enables the seamless amalgamation of models, each specializing in a specific phase density range, thus ensuring a comprehensive and accurate representation of the entire phase space. The MoE model demonstrates its efficacy in accurately capturing complex phase behaviors across various regions of the composition space, as validated by phase diagram comparisons with traditional methods.
	
	Furthermore, the establishment of a high-dimensional phase diagram for the FeCoNiTi system, based on a dense sampling of the phase space, provides a data-rich foundation for future alloy design. The use of the DFS algorithm to delineate independent phase spaces further enhances the utility of this approach for practical applications in materials science.
	
	The methodologies developed in this study lay a robust foundation for the widespread establishment of LCMs across a spectrum of complex alloy systems. This approach heralds a new era in alloy design, moving towards a regularized and efficient high-throughput practice.
	
	\section*{Declaration of Competing Interest}
	The authors declare that they have no known competing financial 
	interests or personal relationships that could have appeared to influence 
	the work reported in this paper.
	
	\section*{Data availability}
	Data will be made available on request.
	
	\section*{Acknowledgments}
	WWS gratefully acknowledges the support of National Natural Science Foundation for Young Scholars of China (Grant No. 52001063), National Natural Science Foundation of China (Grant No. 52371103), and Jiangsu Key Laboratory for Light Metal Alloys (Grant No. LMA202201).
	
	\section*{References}
	\begin{enumerate}
		\item D.B. Miracle, O.N. Senkov, A critical review of high entropy alloys and related concepts, Acta Mater. 122 (2017). https://doi.org/10.1016/j.actamat.2016.08.081.
		\item E. Ghassemali, P.L.J. Conway, High-Throughput CALPHAD: A Powerful Tool Towards Accelerated Metallurgy, Front Mater. 9 (2022). https://doi.org/10.3389/fmats.2022.889771.
		\item N.A.P.K. Kumar, C. Li, K.J. Leonard, H. Bei, S.J. Zinkle, Microstructural stability and mechanical behavior of FeNiMnCr high entropy alloy under ion irradiation, Acta Mater. 113 (2016). https://doi.org/10.1016/j.actamat.2016.05.007.
		\item Y. Shi, B. Yang, P.K. Liaw, Corrosion-resistant high-entropy alloys: A review, Metals (Basel). 7 (2017). https://doi.org/10.3390/met7020043.
		\item X.L. An, Z.D. Liu, L.T. Zhang, Y. Zou, X.J. Xu, C.L. Chu, W. Wei, W.W. Sun, A new strong pearlitic multi-principal element alloy to withstand wear at elevated temperatures, Acta Mater. 227 (2022) 117700. https://doi.org/https://doi.org/10.1016/j.actamat.2022.117700.
		\item Y. Zhang, T. Zuo, Y. Cheng, P.K. Liaw, High-entropy alloys with high saturation magnetization, electrical resistivity, and malleability, Sci Rep. 3 (2013). https://doi.org/10.1038/srep01455.
		\item P. Koželj, S. Vrtnik, A. Jelen, S. Jazbec, Z. Jagličić, S. Maiti, M. Feuerbacher, W. Steurer, J. Dolinšek, Discovery of a superconducting high-entropy alloy, Phys Rev Lett. 113 (2014). https://doi.org/10.1103/PhysRevLett.113.107001.
		\item J.W. Yeh, Alloy design strategies and future trends in high-entropy alloys, JOM. 65 (2013). https://doi.org/10.1007/s11837-013-0761-6.
		\item Y. Zhang, T.T. Zuo, Z. Tang, M.C. Gao, K.A. Dahmen, P.K. Liaw, Z.P. Lu, Microstructures and properties of high-entropy alloys, Prog Mater Sci. 61 (2014). https://doi.org/10.1016/j.pmatsci.2013.10.001.
		\item E.P. George, W.A. Curtin, C.C. Tasan, High entropy alloys: A focused review of mechanical properties and deformation mechanisms, Acta Mater. 188 (2020). https://doi.org/10.1016/j.actamat.2019.12.015.
		\item S. Zhu, J. Shittu, A. Perron, C. Nataraj, J. Berry, J.T. McKeown, A. van de Walle, A. Samanta, Probing phase stability in CrMoNbV using cluster expansion method, CALPHAD calculations and experiments, Acta Mater. 255 (2023). https://doi.org/10.1016/j.actamat.2023.119062.
		\item J. Du, V. Jindal, A.P. Sanders, K.S. Ravi Chandran, CALPHAD-guided alloy design and processing for improved strength and toughness in Titanium Boride (TiB) ceramic alloy containing a ductile phase, Acta Mater. 171 (2019). https://doi.org/10.1016/j.actamat.2019.03.040.
		\item A. van de Walle, H. Chen, H. Liu, C. Nataraj, S. Samanta, S. Zhu, R. Arroyave, Interactive Exploration of High-Dimensional Phase Diagrams, JOM. 74 (2022). https://doi.org/10.1007/s11837-022-05314-z.
		\item L. Zhao, L. Jiang, L.X. Yang, H. Wang, W.Y. Zhang, G.Y. Ji, X. Zhou, W.A. Curtin, X.B. Chen, P.K. Liaw, S.Y. Chen, H.Z. Wang, High throughput synthesis enabled exploration of CoCrFeNi-based high entropy alloys, J Mater Sci Technol. 110 (2022). https://doi.org/10.1016/j.jmst.2021.09.031.
		\item J. Zhang, R. Wang, Y. Zhong, Identification of the Eutectic Points in the Multicomponent Systems with the High-Throughput CALPHAD Approach, J Phase Equilibria Diffus. 43 (2022). https://doi.org/10.1007/s11669-022-01003-1.
		\item A. Abu-Odeh, E. Galvan, T. Kirk, H. Mao, Q. Chen, P. Mason, R. Malak, R. Arróyave, Efficient exploration of the High Entropy Alloy composition-phase space, Acta Mater. 152 (2018). https://doi.org/10.1016/j.actamat.2018.04.012.
		\item J. Gao, J. Zhong, G. Liu, S. Yang, B. Song, L. Zhang, Z. Liu, A machine learning accelerated distributed task management system (Malac-Distmas) and its application in high-throughput CALPHAD computation aiming at efficient alloy design, Advanced Powder Materials. 1 (2022). https://doi.org/10.1016/j.apmate.2021.09.005.
		\item Y. Zeng, M. Man, K. Bai, Y.W. Zhang, Explore the full temperature-composition space of 20 quinary CCAs for FCC and BCC single-phases by an iterative machine learning + CALPHAD method, Acta Mater. 231 (2022). https://doi.org/10.1016/j.actamat.2022.117865.
		\item P. Indyk, R. Motwani, Approximate nearest neighbors: Towards removing the curse of dimensionality, in: Conference Proceedings of the Annual ACM Symposium on Theory of Computing, 1998.
		\item K. pong Chan, A.W. chee Fu, Efficient time series matching by wavelets, Proc Int Conf Data Eng. (1999). https://doi.org/10.1109/icde.1999.754915.
		\item Z. Liu, X. An, W. Sun, The High-dimensional Phase Diagram and the Large CALPHAD Model, (2023). https://arxiv.org/abs/2311.07174.
		\item H. Hosoda, S. Miyazaki, Y. Mishima, Phase constitution of some intermetallics in continuous quaternary pillar phase diagrams, Journal of Phase Equilibria. 22 (2001). https://doi.org/10.1361/105497101770332947.
		\item A.M. Cardinale, N. Parodi, A. Saccone, The 500 °C isothermal section of the Tb–Al–Si system and thermal behavior of selected Al-rich alloys, J Therm Anal Calorim. 130 (2017). https://doi.org/10.1007/s10973-017-6241-4.
		\item A.J. Zaddach, Zaddach, A. Joseph, Physical Properties of NiFeCrCo-based High-Entropy Alloys, PhDT. (2015).
		\item L. Zhao, L. Jiang, L.X. Yang, H. Wang, W.Y. Zhang, G.Y. Ji, X. Zhou, W.A. Curtin, X.B. Chen, P.K. Liaw, S.Y. Chen, H.Z. Wang, High throughput synthesis enabled exploration of CoCrFeNi-based high entropy alloys, J Mater Sci Technol. 110 (2022). https://doi.org/10.1016/j.jmst.2021.09.031.
		\item R.A. Jacobs, M.I. Jordan, S.J. Nowlan, G.E. Hinton, Adaptive Mixtures of Local Experts, Neural Comput. 3 (1991). https://doi.org/10.1162/neco.1991.3.1.79.
		\item M.J. Sippl, H.A. Scheraga, Cayley-Menger coordinates., Proc Natl Acad Sci U S A. 83 (1986). https://doi.org/10.1073/pnas.83.8.2283.
		
	\end{enumerate}

\end{document}